\documentclass{article}
\usepackage[utf8]{inputenc}
\usepackage{authblk}
\usepackage{setspace}
\usepackage[margin=1.25in]{geometry}
\usepackage{graphicx}
\graphicspath{ {./figures/} }
\usepackage{subcaption}
\usepackage{amsmath}
\usepackage{lineno}
\usepackage{float}
\usepackage{hyperref}
\usepackage{engord}
\usepackage{booktabs}
\usepackage{soul}
\usepackage{fancyhdr}
\usepackage{multirow}
\usepackage{pdflscape}
\usepackage{tikz}
\usepackage{pgfmath}
\usepackage{tikz}
\usepackage{quantikz}
\usepackage{xr-hyper} 
\usepackage{hyperref} 

\definecolor{initpurple}{RGB}{155, 30, 75}
\definecolor{hred}{RGB}{235, 65, 65}

\tikzset{
    init/.style={fill=white, draw=black, text=white, font=\tiny, inner sep=2pt, rounded corners=1pt},
    hstyle/.style={fill=white, draw=black, font=\sffamily\scriptsize, inner sep=2pt, rounded corners=1pt}
}

\definecolor{myblue}{RGB}{15, 35, 90}
\definecolor{edgegrey}{RGB}{160,160,160}

\newcommand{\rowlabel}[1]{{\footnotesize #1}}

\usepackage[table]{xcolor}
\usepackage{pgfplots}
\pgfplotsset{compat=1.17}

\newcommand{\colcell}[1]{%
    \pgfmathparse{#1*100}%
    \edef\temp{\noexpand\cellcolor{green!\pgfmathresult!yellow}}%
    \temp #1%
}

\engordraisetrue

\usepackage[bottom]{footmisc}

\usepackage{tikz}
\usetikzlibrary{arrows.meta,positioning,fit,backgrounds,calc}

\usepackage[style=phys, citestyle=numeric-comp, sorting=none, backend=biber, eprint=true]{biblatex}
\title{Benchmarking Quantum Computers via Protocols\\Comparing Superconducting and Ion-Trap Quantum Technology}

\author{$^*$Nitay Mayo, $^{*\dagger}$Tal Mor and $^*$Yossi Weinstein}
\affil{$^*$Department of Computer Science, Technion University, Haifa, Israel.}
\affil{$^\dagger$The Helen Diller Quantum Center}
\date{\today}
\begin{document}

\maketitle

\begin{abstract}
Both Superconducting and Ion-Trap are leading quantum architectures common in the current landscape of the quantum computing field, each with distinct characteristics and operational constraints. Understanding and measuring the underlying \underline{quantumness} of these devices is essential for assessing their readiness for practical applications and guiding future progress and research. Building on earlier work (Meirom, Mor and Weinstein Arxiv 2505.12441), we utilize a benchmarking strategy applicable for comparing these two architectures by measuring "quantumness" directly on optimal sub-chips. Distinct from existing metrics, our approach employs rigorous binary fidelity thresholds derived from the classical limits of state transfer. This enables us to definitively establish quantum advantage of a designated sub-region. Here we apply this quality assurance methodology to platforms from both technologies. This comparison provides a protocol-based evaluation of quantumness advantage, revealing not only the strengths and weaknesses of each tested chip and its sub-chips but also offering a common language for their assessment. By abstracting away technical differences in the final result, we demonstrate a benchmarking strategy that bridges the gap between disparate quantum-circuit technologies, enabling fair performance comparisons and establishing a critical foundation for evaluating future claims of quantum advantage.
We encourage other companies to emulate the independent qubit availability and the fair pricing, as do IBM and AQT, which allow researchers to perform such assessments. 
In the discussion section, we consider an important future direction of using "benchmarking via protocols" also for fault-tolerant quantum computing with logical qubits.

\end{abstract}
 \newpage
\tableofcontents

\section{Introduction}
The characterization of Quantum Computers (QC) in the ``Noisy Intermediate-Scale Quantum'' (NISQ) era \cite{Preskill_2018} —highlighted by recent milestone demonstrations of quantum supremacy~\cite{Arute_2019}, evidence of utility before fault tolerance~\cite{kim2023evidence}, the advent of logical quantum processors on atom arrays~\cite{Bluvstein_2023, xu2024constant}, and the experimental demonstration of logical magic state distillation~\cite{Bravyi_2005, meier_2013, rodriguez_2025} — necessitates robust benchmarking frameworks that transcend simple gate fidelities. Early foundational methods, such as Randomized Benchmarking (RB) \cite{Emerson_2005, PhysRevA.77.012307}, focused on evaluating the average error rates across random unitary operations. These have evolved into volumetric frameworks like Quantum Volume (QV)~\cite{Cross_2019, Blume_Kohout_2020}, developed by IBM, QV provides a consolidated, single-number metric for effective computational power by evaluating a system's performance on randomized square circuits—those where circuit depth and width are equal. Recent demonstrations have pushed these metrics further, such as achieving a QV of 64 on superconducting systems \cite{Jurcevic_2021}. While these volumetric metrics are useful for high-level comparisons, they often lack the granularity needed to diagnose specific architectural bottlenecks or sub-region performance \cite{Proctor_2021}.

In this work we apply a novel benchmarking method~\cite{Bench1_arxiv} to two distinct types of quantum computers with significant architectural differences, thereby illustrating the utility of our benchmarking approach for comparative performance analysis across diverse platforms. To demonstrate quantum behavior of a chip, we utilize a benchmarking strategy based on quantum communication and state transfer primitives~\cite{PhysRevLett.70.1895, PhysRevLett.69.2881}. The method relies on a binary assessment of ``quantumness'', where success is defined by exceeding fidelity thresholds that have been formally established as the classical limit for degraded qubit teleportation~\cite{PhysRevLett.74.1259} and for degraded shared fully entangle state~\cite{PhysRevA.40.4277} such as the singlet state. This methodology was previously borrowed~\cite{Bench1_arxiv} for the degradation of a transferred qubit and transferred singlet state, and the methodology was later applied to compare internal generations of superconducting hardware~\cite{Bench2_arxiv}, specifically the IBM Heron and Eagle processors. In this paper, in addition to comparing two distinct QC architectures, we extend this body of work by introducing the \textit{Generalized Transmit} protocol and adapting the Optimal Lookup Workflow to bridge the gap between ion-trap and superconducting architectures.


Reflecting these diverse topologies, our study first evaluates AQT's IBEX Q1, a 12-qubit quantum computer that utilizes trapped-ion technology. The all-to-all connectivity of IBEX Q1, mediated by a collective phonon mode, implies that a quantum operation can be executed between any arbitrary pair of qubits without intermediate routing.
The comparative analysis includes two superconducting systems from IBM: Fez, a 156-qubit processor from the Heron-r2 series, and Brisbane, a 127-qubit processor from the Eagle-r3 series. Both superconducting devices feature qubits arranged in a fixed rectangular lattice. Although Fez and Brisbane possess significantly more qubits than IBEX Q1, their connectivity constraints are stricter; only near-neighboring qubits can interact through quantum gates.

Our benchmarking method is based on identifying quantumness on a protocol-level performance. A series of three protocols and their generalized versions are applied on each quantum computer according to an assessment methodology tailored to the specific architecture. At the conclusion of this process, our methodology identifies sub-regions that are considered optimal for a specific protocol, as detailed in Section~\ref{sec:optimal_lookup_workflow}. A key strength of our approach is that the performance of these optimal sub-regions can be directly compared, regardless of the underlying architecture. 


The remainder of this paper is organized as follows: \\
Section~\ref{sec:protocols_explanations} contains information about the protocols used in this work while also introducing a novel one. Section~\ref{sec:optimal_lookup_workflow} defines the assessment methodology - the "Optimal Lookup Workflow" used to assess the three QCs. Section~\ref{sec:hardware_intro} supplies some information and connectivity maps of the three evaluated QCs. In Section~\ref{sec:comparison_section} we briefly present the results of the experiments and compare the performance of the three QCs, along with a discussion on their advantages and limitations. Sub-section~\ref{sec:limitations_and_malfunctions} discusses the limitations and malfunctions we encountered during this research. Sub-section~\ref{sec:benchmarking_modular_qpu} discusses the application of our benchmarking framework to modular quantum architectures and multi-chip interconnects containing 2-3 ``sub-chips'', and in Sub-section~\ref{sec:benchmarking_logical_qpu}, we conduct a short discussion about benchmarking logical QCs. Finally, in Section~\ref{sec:conclusions}, we conclude the research. The full results and charts from the data acquisition process for the three QCs are available in the corresponding supplemental sections~\ref{S-sec:ibex_results_section},~\ref{S-sec:Brisbane_results}, and \ref{S-sec:fez_results_section}. 


\subsection{Related Work and Current Contribution}

The evaluation of QPUs in the NISQ era has driven the development of comprehensive benchmarking suites and application-oriented strategies \cite{Lubinski_2023}. Frameworks such as QASMBench \cite{Ang_2023} and MQT Bench \cite{Quetschlich_2023} provide extensive datasets based on foundational quantum procedures (e.g., the Deutsch-Jozsa algorithm, Grover's algorithm, QFT, and QPE), while other approaches evaluate performance using specific algorithms like VQE or QAOA. Direct cross-platform studies, such as the comparison between trapped-ion and superconducting architectures by Linke et al. \cite{Linke_2017}, highlight how fundamental hardware characteristics critically impact these benchmarking procedures. For instance, trapped-ion systems typically offer all-to-all connectivity via shared phonon modes \cite{Bruzewicz_2019, PhysRevLett.74.4091}, whereas superconducting systems are generally restricted to fixed nearest-neighbor rectangular lattices \cite{Jurcevic_2021}. Despite these advancements, current application-based methodologies face notable limitations. The deep circuits they generate are often too demanding for the limited coherence times and error rates of existing hardware. Furthermore, they primarily focus on tool-agnostic comparative evaluations—measuring metrics like compilation runtime or gate count reductions relative to a baseline—rather than establishing rigorous success or failure criteria, ultimately lacking definitive thresholds to certify quantum performance.

In parallel, the severe performance degradation caused by SWAP gate overhead in restricted hardware topologies has shown the importance of hardware-aware circuit compilation \cite{10.1145/3297858.3304023} and qubits mapping. Subsequent research has reinforced this by proposing spatial-aware routing heuristics and noise-adaptive compiler mappings that dynamically assign logical to physical qubits based on the specific connectivity and calibration data of the hardware \cite{murali2019noiseadaptivecompilermappingsnoisy, Niu_2020}. Together, these works emphasize the need for a connectivity-aware, while also technology-agnostic, benchmarking strategy capable of isolating and evaluating high-performing sub-regions within a given chip.

Addressing the gaps in existing evaluation methods, the primary strength of our approach is the utilization of a definitive success criterion that provides formal evidence of above-classical performance~\cite{Bench1_arxiv, Bench2_arxiv}. Unlike volumetric or algorithmic benchmarks, our protocol-based approach establishes rigorous fidelity thresholds derived from classical limits, enabling a definitive demonstration of quantum advantage at the sub-chip level. Furthermore, another noted contribution presented in this work is the template "Optimal Lookup Workflow" designed to systematically segment a QPU, regardless of the implementation technology or connectivity constraints, and identify its best-performing regions. The output of this benchmarking methodology yields valuable spatial performance data, which can be fed directly into a spatial-aware compiler to achieve significant performance enhancements in the fidelity of the resulting compiled circuit. 


\section{The Protocols}\label{sec:protocols_explanations}
In this study, we employ three protocols alongside their generalized versions for two of them. Architectural differences in qubit connectivity necessitated specific adjustments to the protocol definitions for each platform. This section details these adaptations and provides a formal definition for the ``\textit{Generalized Transmit}'' protocol, developed specifically for this work.

A primary differentiator between the architectures in the context of protocol performance is the ``Swap Distance'', a parameter used in previous work~\cite{Bench1_arxiv, Bench2_arxiv}. Swap distance is defined as the number of swap-gates used to move the state of the work qubits between Alice and Bob in the protocol. 

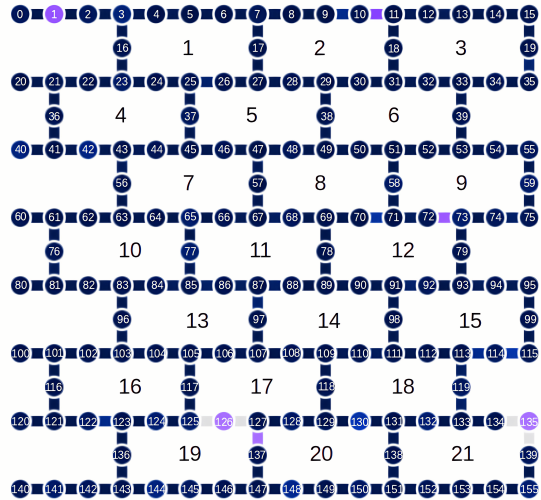
\begin{figure}[htpb]
    \centering
    
    \begin{tikzpicture}[
        x=0.95cm, y=0.8cm,
        nnode/.style={circle, fill=myblue, text=white, inner sep=0pt, minimum size=20pt, font=\sffamily\small\bfseries},
        edge/.style={line width=4.5pt, edgegrey}
    ]
    
    \foreach \r in {0,...,7} {
        \draw[edge] (0, -\r*2) -- (15, -\r*2);
    }
    
    \foreach \r in {0,2,4,6} {
        \foreach \c in {3, 7, 11, 15} {
            \draw[edge] (\c, -\r*2) -- (\c, -\r*2-2);
        }
    }
    \foreach \r in {1,3,5} {
        \foreach \c in {1, 5, 9, 13} {
            \draw[edge] (\c, -\r*2) -- (\c, -\r*2-2);
        }
    }
    
    \draw[edge] (0, 0) -- (2, 0);       
    \draw[edge] (10, 0) -- (12, 0);     
    \draw[edge] (12, -6) -- (14, -6);   
    \draw[edge] (5, -12) -- (7, -12);   
    \draw[edge] (14, -12) -- (15, -12); 
    
    \foreach \r in {0,...,7} {
        \foreach \c in {0,...,15} {
            \pgfmathsetmacro{\val}{int(\r*20 + \c)}
            
            \ifnum\val=1 \def\nstyle{nnode}
            \else\ifnum\val=11 \def\nstyle{nnode}
            \else\ifnum\val=73 \def\nstyle{nnode}
            \else\ifnum\val=126 \def\nstyle{nnode}
            \else\ifnum\val=135 \def\nstyle{nnode}
            \else \def\nstyle{nnode} \fi\fi\fi\fi\fi
            
            \node[\nstyle] at (\c, -\r*2) {\val};
        }
    }
    
    \foreach \r in {0,2,4,6} {
        \foreach \i [count=\xi from 0] in {3, 7, 11, 15} {
            \pgfmathsetmacro{\val}{int(\r*20 + 16 + \xi)}
            \node[nnode] at (\i, -\r*2 - 1) {\val};
        }
    }
    \foreach \r in {1,3,5} {
        \foreach \i [count=\xi from 0] in {1, 5, 9, 13} {
            \pgfmathsetmacro{\val}{int(\r*20 + 16 + \xi)}
            \node[nnode] at (\i, -\r*2 - 1) {\val};
        }
    }
    
    \foreach \i in {1,2,3} {
        \node[font=\sffamily\Huge] at (1+4*\i, -1) {\i};
        \pgfmathsetmacro{\v}{int(\i+6)}
        \node[font=\sffamily\Huge] at (1+4*\i, -5) {\v};
        \pgfmathsetmacro{\v}{int(\i+12)}
        \node[font=\sffamily\Huge] at (1+4*\i, -9) {\v};
        \pgfmathsetmacro{\v}{int(\i+18)}
        \node[font=\sffamily\Huge] at (1+4*\i, -13) {\v};
    }
    \foreach \i in {1,2,3} {
    \pgfmathsetmacro{\v}{int(\i+3)}
    \node[font=\sffamily\Huge] at (-1+4*\i, -3) {\v};
    \pgfmathsetmacro{\v}{int(\i+9)}
    \node[font=\sffamily\Huge] at (-1+4*\i, -7) {\v};
    \pgfmathsetmacro{\v}{int(\i+15)}
    \node[font=\sffamily\Huge] at (-1+4*\i, -11) {\v};
}
    
    \end{tikzpicture}
    
    \caption{IBM's Heron-r2 series qubits connectivity map. Each rectangle is assigned a number for consistent reference}
    \label{fig:fez_qubit_map}
\end{figure}

In the context of Ion-Trap architecture, the basic protocols utilized in~\cite{Bench2_arxiv} do not sufficiently characterize a practical reduced chip. For example, in IBEX Q1 the only possible swap distance between any two qubits is one, as seen in the left side of Figure~\ref{fig:ibex_qubit_map}, because every qubit is connected to every other qubit. For that reason, generalized protocols were defined in~\cite{Bench1_arxiv}, protocols that provide a fairer comparison between these two architectures.
\begin{figure}[htbp]
    \centering


    \begin{tikzpicture}[
    node_style/.style={circle, draw=black, fill=myblue, text=white, minimum size=0.6cm, inner sep=0pt, font=\sffamily\small}, edge_style/.style={edgegrey, line width=0.8pt}
]

    \begin{scope}[name prefix=G1-]
        \foreach \i in {0,...,11} {
            \node[node_style] (n\i) at ({30*\i}:2.5) {\i};
        }
        
        \foreach \i in {0,...,10} {
            \foreach \j in {0,...,11} {
                \pgfmathtruncatemacro{\jpos}{\j}
                \ifnum\i<\jpos
                    \draw[edge_style] (n\i) -- (n\j);
                \fi
            }
        }
    \end{scope}

    \begin{scope}[xshift=7cm, name prefix=G2-]
        \node[node_style] (P) at (0,0) {P};
        
        \foreach \i in {0,...,11} {
            \node[node_style] (n\i) at ({30*\i}:2.5) {\i};
        }
        
        \foreach \i in {0,...,11} {
            \draw[edge_style] (P) -- (n\i);
        }
    \end{scope}

\end{tikzpicture}

    \caption{Connectivity topology of the AQT IBEX Q1 processor. (a) The effective logical connectivity, illustrated as a complete graph. The system supports all-to-all connectivity, allowing the execution of arbitrary two-qubit gates between any pair of the 12 ions without intermediate SWAP operations. (b) We speculate that the physical interaction mechanism is actually a star topology as originally suggested by Cirac and Zoller \cite{PhysRevLett.74.4091}. Connectivity is mediated by a centralized, shared phonon mode (P). While this common bus facilitates the all-to-all coupling seen in (a), it necessitates making the entangling operations in a series instead of in parallel. The central phonon mode acts solely as a mediator and is not part of the computational qubits set. Note that this graph is only an illustration. Physically the ions lay in line and not in a circle}
            \label{fig:ibex_qubit_map}    

\end{figure}
In contrast to IBEX Q1, the fixed rectangular lattice topology of the IBM processors necessitates routing chains for non-adjacent qubits. By partitioning the IBM chips into rectangular sub-regions (as illustrated in Figure~\ref{fig:fez_qubit_map}), we observe the swap distances ranging from one to six gates, depending on the spatial separation within the sub-chip.


We focus on three primary protocols: \textit{Transmit}, \textit{Do-nothing}, and \textit{Bell-state transfer}. These were selected for their fundamental nature and the existence of straightforward generalized variants. Consistent with the definitions in~\cite{Bench2_arxiv, Bench1_arxiv, PhysRevLett.74.1259}, we establish a fidelity threshold of 2/3 for both the \textit{transmit} and \textit{do-nothing} protocols. For the \textit{Bell-state transfer}, which relies on entanglement, the threshold is set at 0.5 following~\cite{Bench1_arxiv, Bench2_arxiv, PhysRevA.40.4277}. These thresholds were used by others in the past to prove a successful teleportation~\cite{Pfaff_2014} or a successful entangled state creation~\cite{Peters_2004, kwiat_2004, zeilinger_2012}; later, Meirom, Mor and Weinstein~\cite{Bench1_arxiv} were inspired to adopt these boundaries as quantumness thresholds, as used in this work. For the generalized version of \textit{transmit} and \textit{do-nothing} protocols, we established a fidelity threshold of $(\frac{2}{3})^M$, where M represents the number of qubits comprising the work state. We could have demanded a more strict threshold of $\frac{2}{3}$ per \emph{each} qubit, however we decided to relax the threshold in this work. While the \textit{Cat state} protocol, defined in~\cite{Bench1_arxiv}, was originally included in the experimental design, it was subsequently and unfortunately excluded due to the budgetary constraints detailed in Section~\ref{sec:limitations_and_malfunctions}.

\begin{figure}[htbp]
    \centering    
    \begin{minipage}{0.42\textwidth}
        \centering 
        \textbf{\Large Ion-Trap}
    \end{minipage}
    \hfill
    \begin{minipage}{0.1\textwidth}
        \centering 
        \textbf{Protocol}
    \end{minipage}
    \hfill
    \begin{minipage}{0.42\textwidth}
        \centering 
        \textbf{\Large Superconducting}
    \end{minipage}

    \begin{subfigure}[c]{0.42\textwidth}
        \includegraphics[width=\linewidth]{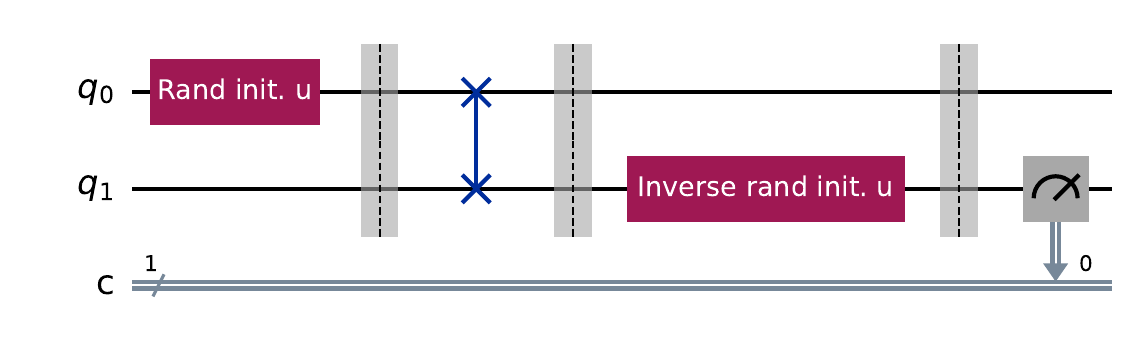}
    \end{subfigure}
    \hfill
    \begin{minipage}[c]{0.1\textwidth}
        \centering 
        \rowlabel{Transmit}
    \end{minipage}
    \hfill
    \begin{subfigure}[c]{0.42\textwidth}
        \includegraphics[width=\linewidth]{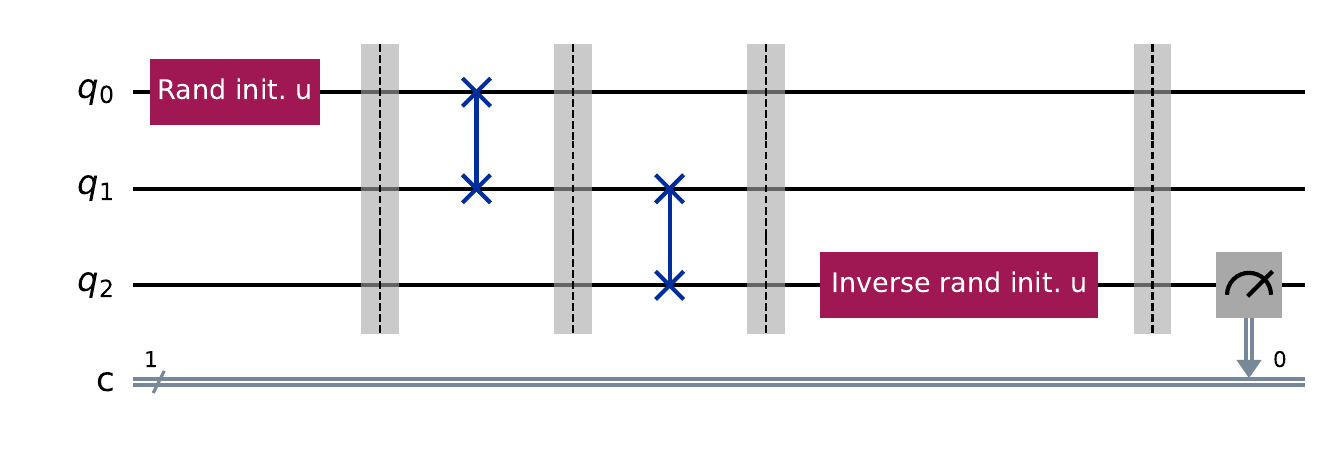}
    \end{subfigure}

    \begin{subfigure}[c]{0.42\textwidth}
        \includegraphics[width=\linewidth]{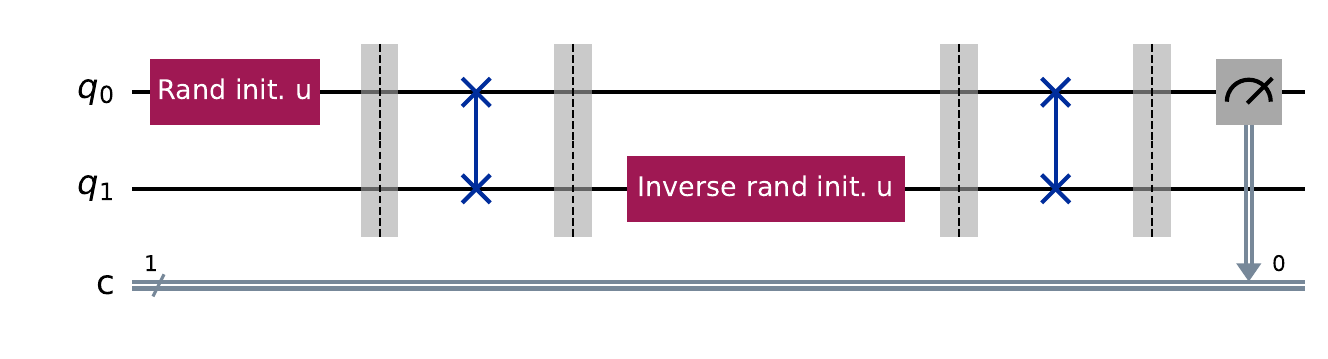}
    \end{subfigure}
    \hfill
    \begin{minipage}[c]{0.1\textwidth}
        \centering 
        \rowlabel{Do-nothing}
    \end{minipage}
    \hfill
    \begin{subfigure}[c]{0.42\textwidth}
        \includegraphics[width=\linewidth]{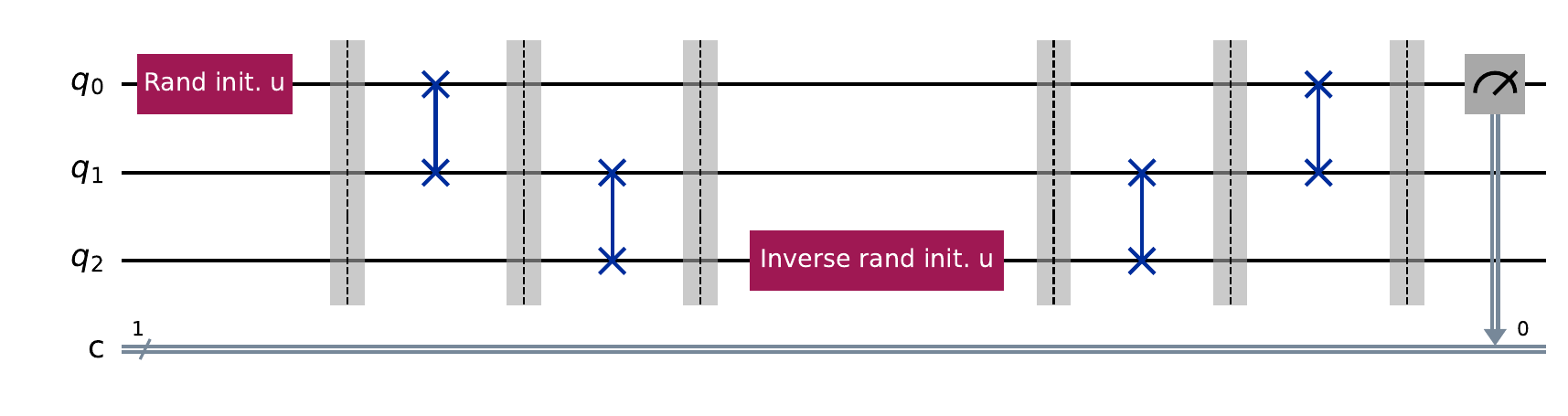}
    \end{subfigure}
    
    \begin{subfigure}[c]{0.42\textwidth}
        \includegraphics[width=\linewidth]{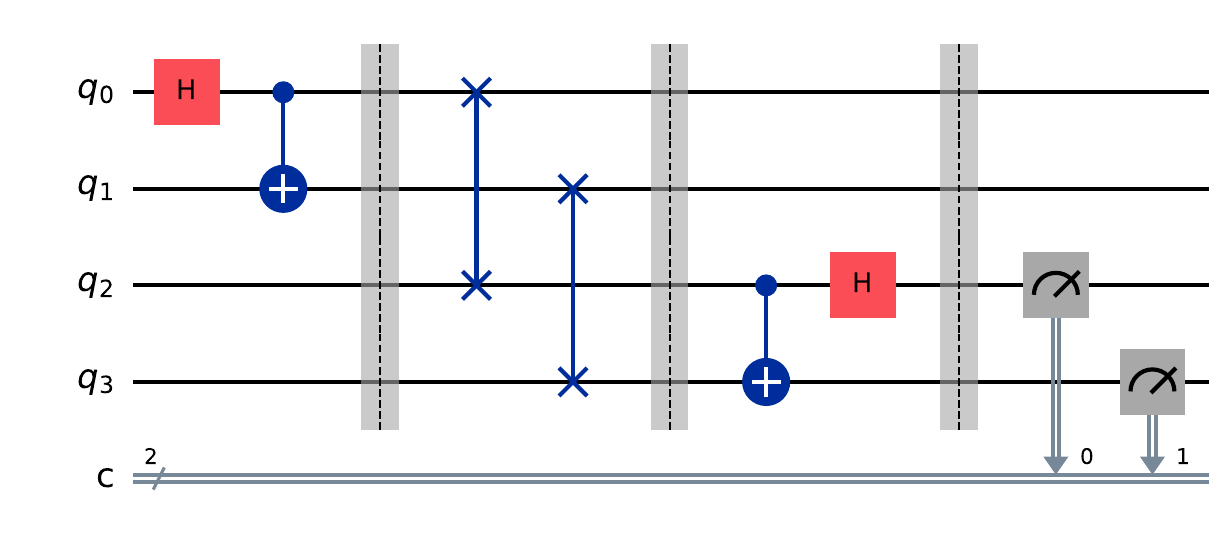}
    \end{subfigure}
    \hfill
    \begin{minipage}[c]{0.1\textwidth}
        \centering 
        \rowlabel{Bell-state transfer}
    \end{minipage}
    \hfill
    \begin{subfigure}[c]{0.42\textwidth}
        \includegraphics[width=\linewidth]{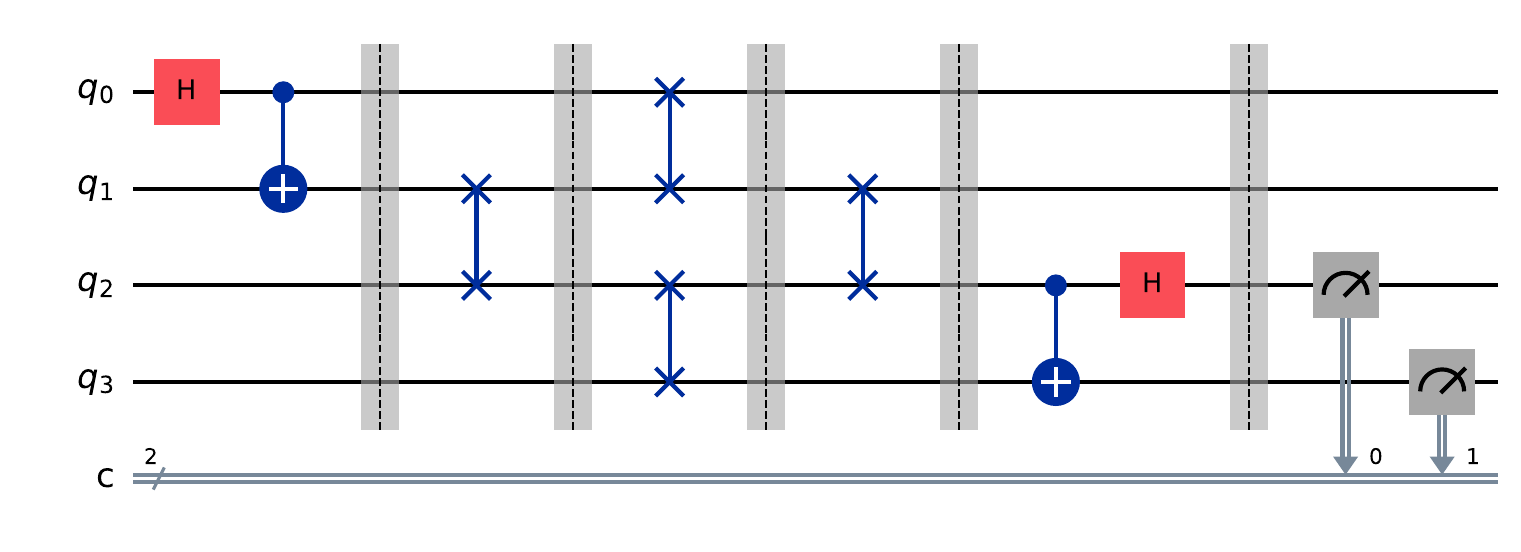}
    \end{subfigure}

    \par\vspace{1em}
    
    \begin{subfigure}[c]{0.42\textwidth}
        \includegraphics[width=\linewidth]{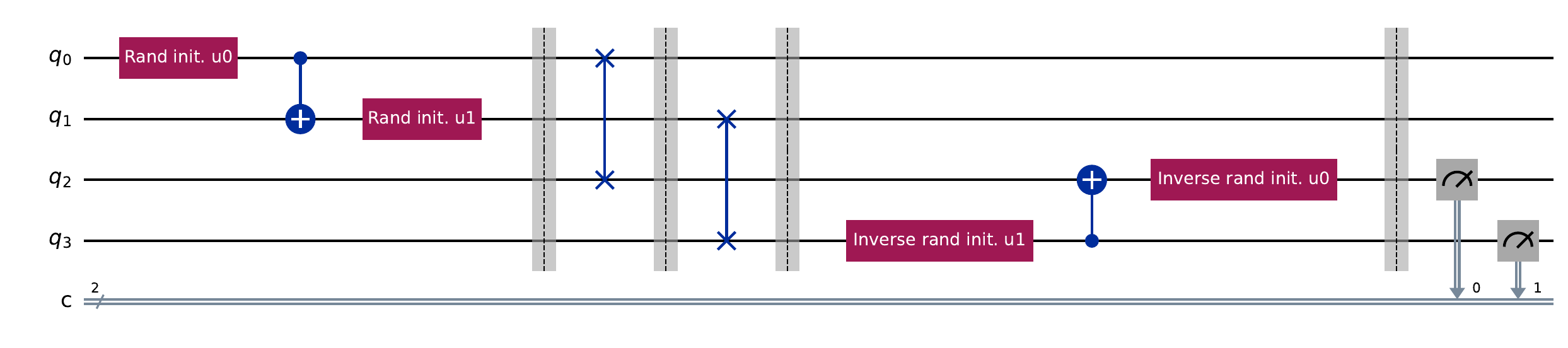}
    \end{subfigure}
    \hfill
    \begin{minipage}[c]{0.1\textwidth}
        \centering 
        \rowlabel{Generalized transmit}
    \end{minipage}
    \hfill
    \begin{subfigure}[c]{0.42\textwidth}
        \includegraphics[width=\linewidth]{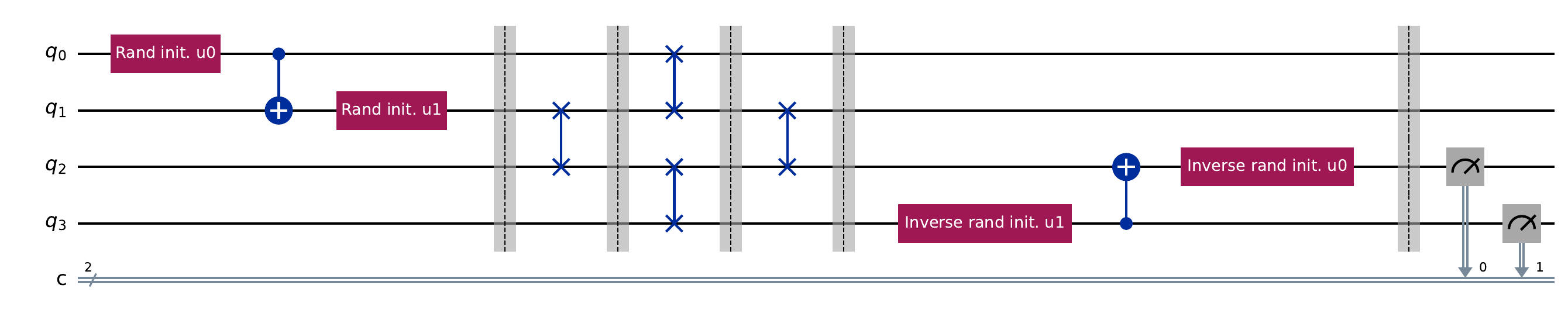}
    \end{subfigure}

    \par\vspace{1em}
    
    \begin{subfigure}[c]{0.42\textwidth}
        \includegraphics[width=\linewidth]{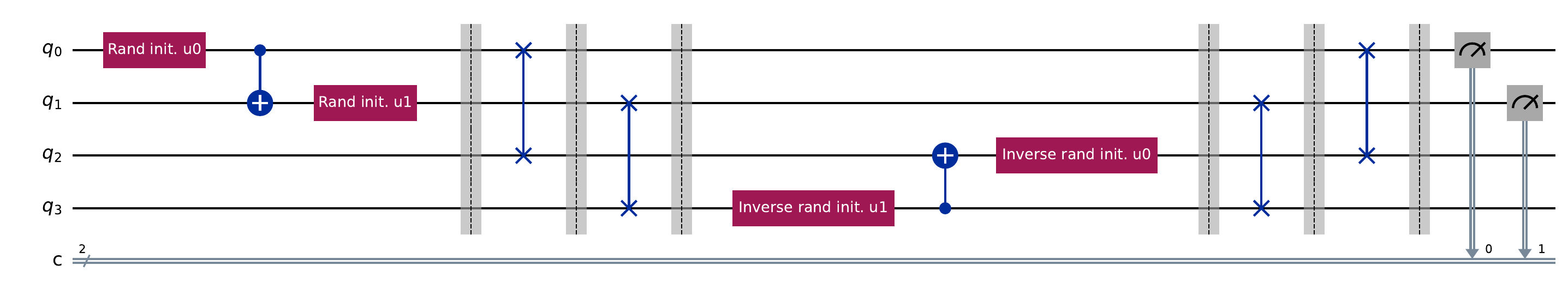}
    \end{subfigure}
    \hfill
    \begin{minipage}[c]{0.12\textwidth}
        \centering 
        \rowlabel{Generalized do-nothing}
    \end{minipage}
    \hfill
    \begin{subfigure}[c]{0.42\textwidth}
        \includegraphics[width=\linewidth]{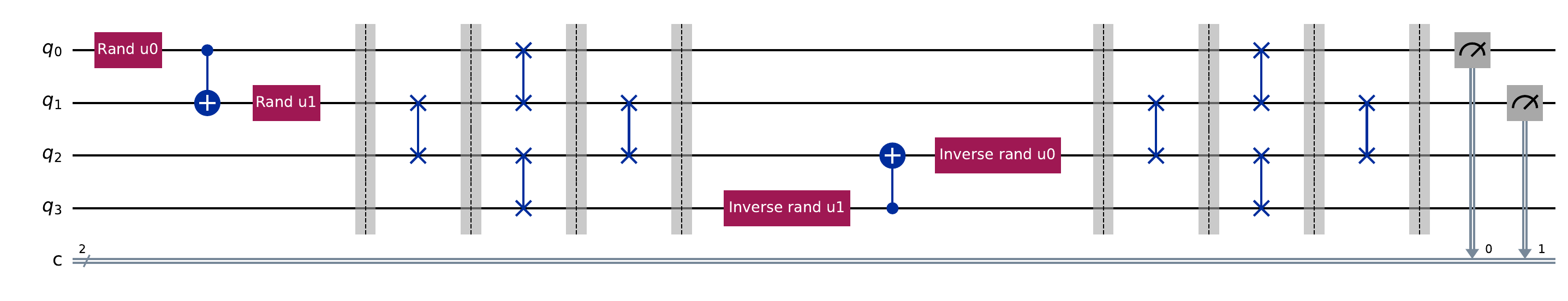}
    \end{subfigure}

    \par\vspace{1em}
    
    \begin{subfigure}[c]{0.42\textwidth}
        \includegraphics[width=\linewidth]{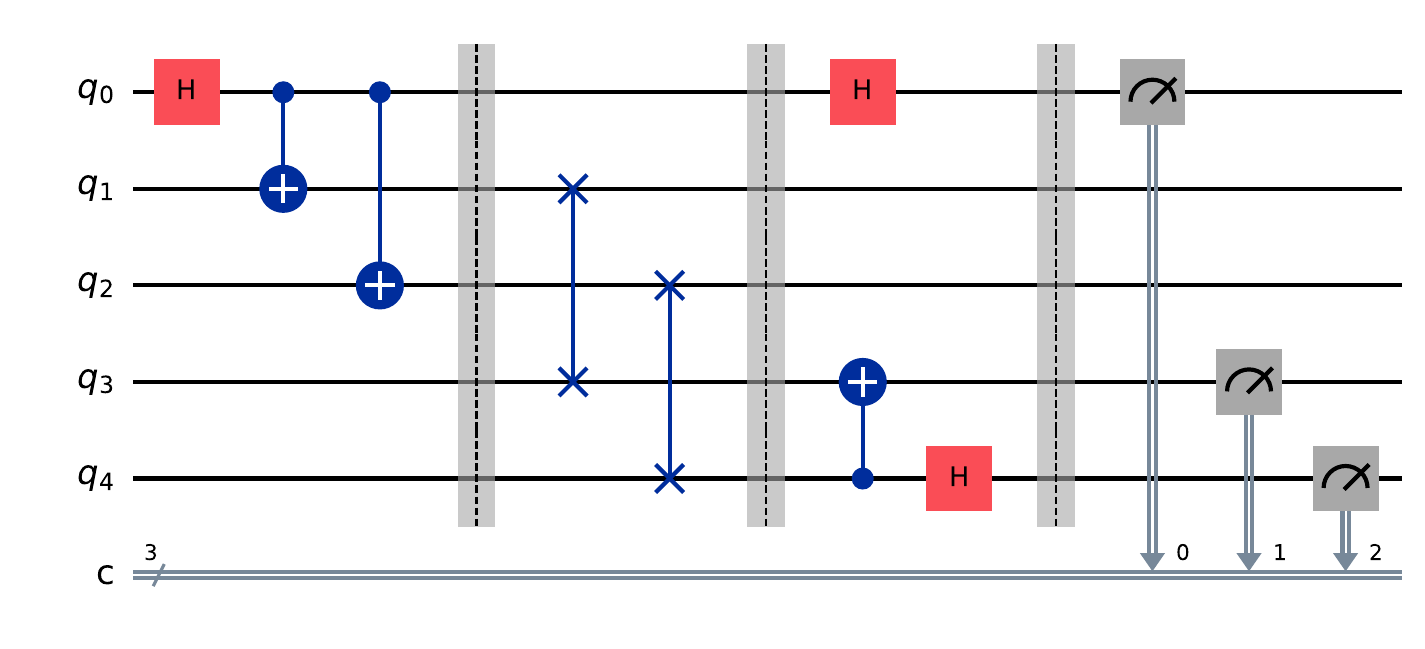}
    \end{subfigure}
    \hfill
    \begin{minipage}[c]{0.12\textwidth}
        \centering 
        \rowlabel{Cat state}
    \end{minipage}
    \hfill
    \begin{subfigure}[c]{0.42\textwidth}
        \includegraphics[width=\linewidth]{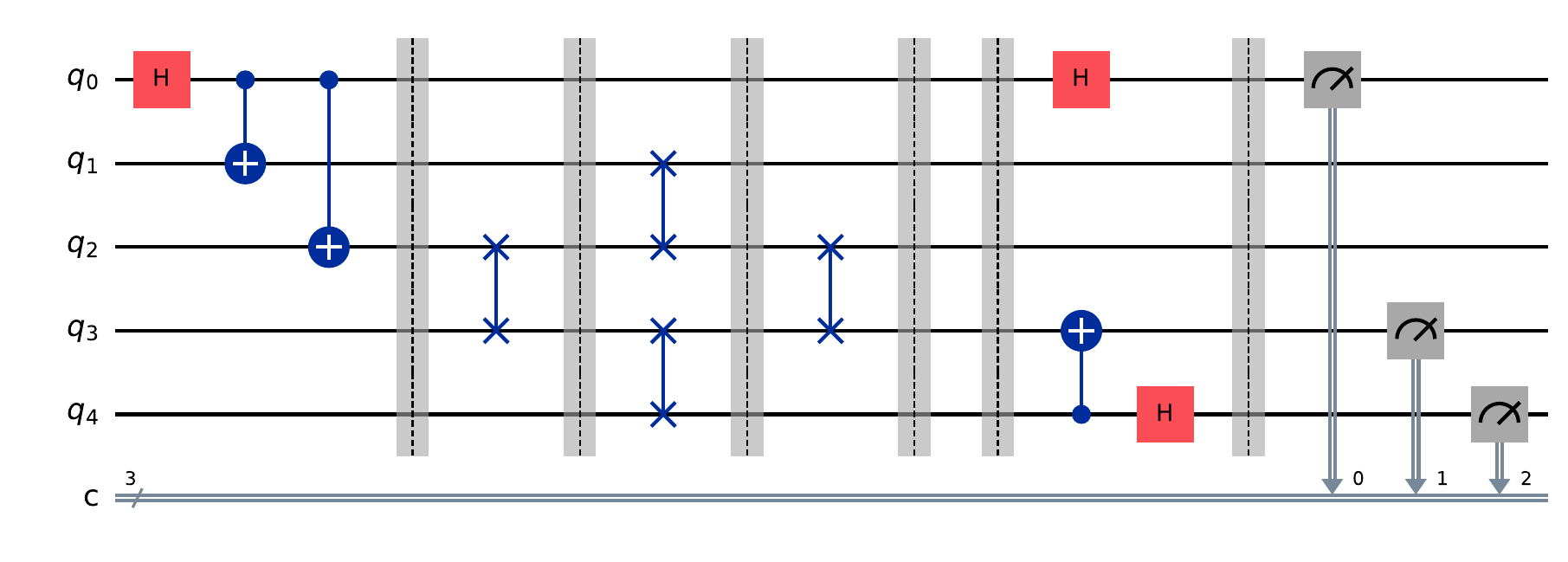}
    \end{subfigure}

    \caption{The protocols used in this work, we adjusted them in order be compatible with AQT's all-to-all connectivity. The \textit{generalized transmit} and \textit{generalized do-nothing} protocols are presented with $M=2$. On AQT's quantum computer the swap distance between 2 qubits is always 1, while in IBM's hardware the swap distance depends on the path we choose}
    \label{fig:adjusted_protocols}
\end{figure}

Figure~\ref{fig:adjusted_protocols} shows a comparison between the versions of the protocols that are applied to each quantum computer, including the \textit{cat state} protocol which as discussed, does not take part in the comparison presented in this work.

\subsection{A Note About The Protocols Transmit and Generalized Transmit}
The \textit{transmit} protocol was initially defined in~\cite{Bench2_arxiv} to accommodate the hardware limitations encountered during the preliminary stages of this research. Its simplicity proved essential for refining the comparative analysis of lower-performing hardware. \textit{Generalized transmit} is a novel protocol added here, it is a natural extension of \textit{transmit}. When applying the \textit{generalized transmit} protocol we set a parameter M as the number of qubits in the transmitted state as was also done for \textit{do-nothing} protocol in~\cite{Bench1_arxiv}. Figure~\ref{fig:adjusted_protocols} illustrates a \textit{generalized transmit} protocol with $M=2$. 
The parameter M is also relevant when applying the \textit{generalized do-nothing} protocol and serves the same purpose: determining the number of qubits in the work qubits state.

\subsection{Identifying Good Quantum Sub-Chips}
A primary output of our benchmarking methodology is the characterization of an ``Optimal Sub-Chip''. We define this as the largest contiguous sub-region of a processor that demonstrated a consistent quantumness advantage across the tested protocols. The identification of these regions is a natural result of the Optimal Lookup Workflow, which is detailed in Section~\ref{sec:optimal_lookup_workflow}. For the end user, this metric is critical; it serves as a realistic indicator of the system's effective computational size and identifies the specific physical qubits where quantum performance is most reliable.

The physical implementation of an optimal sub-chip is dictated by the specific connectivity constraints of the architecture. For IBM's processors, the optimal sub-chip represents the maximal set of neighboring 12-qubit rectangles that successfully pass the full assessment for all protocols; in cases where no such unified region exists, it is defined as the maximal set of rectangles that were found to be optimal for any individual protocol within the workflow. 

Conversely, the all-to-all connectivity of the ion-trap architecture necessitates a different approach where the optimal sub-chip, or ``Reduced Chip", is defined as the maximal set of individual qubits that demonstrate quantum advantage for any individual protocol. We note that unlike the fixed rectangular granularity of the IBM systems, the ion-trap's optimal sub-chip is defined per protocol to accurately reflect its unique logical topology and the results of the adapted benchmarking methodology.

\section{The Optimal Lookup Workflow}\label{sec:optimal_lookup_workflow}
Consistent with previous work, the comparison of the two quantum computers begins with individual assessment of each. However, the optimal lookup workflow proposed in prior work~\cite{Bench2_arxiv} is incompatible with the specific connectivity constraints of AQT's quantum computer architecture. Consequently, the assessment process was adapted. In AQT's optimal lookup workflow we apply a series of experiments, utilizing first the basic protocols followed by their generalized versions, to generate a comprehensive ``Protocol Vector" characterizing AQT's quantum computer capabilities.\\

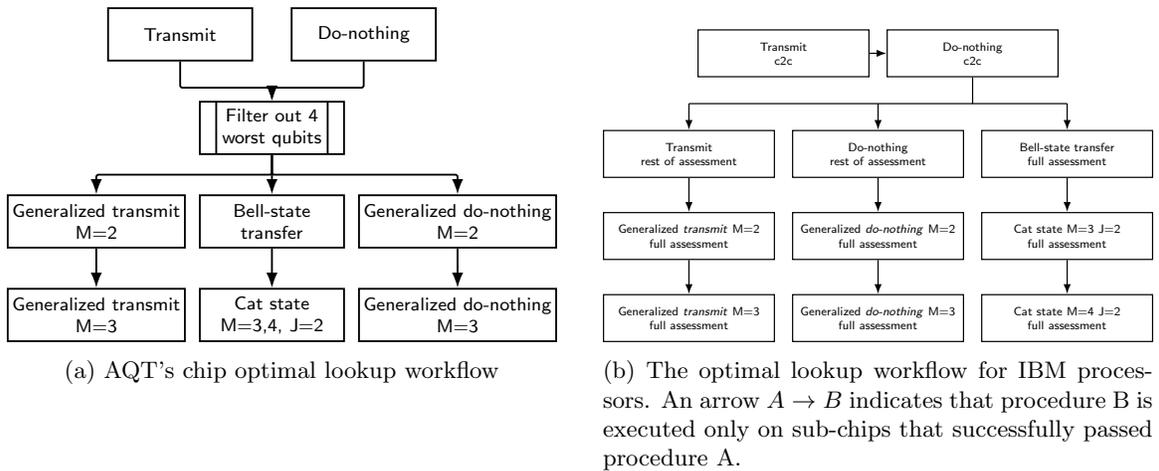
\begin{figure}[htbp]
    \centering

    \begin{subfigure}[t]{0.48\linewidth}
        \centering
    \resizebox{1\textwidth}{!}{%
    \begin{tikzpicture}[
        node distance=0.6cm and 0.2cm,
        font=\footnotesize\sffamily,
        box/.style={
            draw, thick, rectangle, align=center,
            minimum width=2.2cm, minimum height=0.8cm,
            inner sep=2pt
        },
        doublebox/.style={
            box,
            path picture={ 
                \draw[thick] ([xshift=0.25cm]path picture bounding box.north west) -- ([xshift=0.25cm]path picture bounding box.south west);
                \draw[thick] ([xshift=-0.25cm]path picture bounding box.north east) -- ([xshift=-0.25cm]path picture bounding box.south east);
            }
        },
        line/.style={thick, -{Latex}, rounded corners=1pt}
    ]

        \node[doublebox] (filter) {Filter out 4\\worst qubits};
        
        \node[box, above=of filter, xshift=-1.4cm] (transmit) {Transmit};
        \node[box, above=of filter, xshift=1.4cm] (donothing) {Do-nothing};
        
        \node[box, below=of filter] (bell) {Bell-state\\transfer};
        \node[box, left=0.2cm of bell] (gen_trans_2) {Generalized transmit\\M=2};
        \node[box, right=0.2cm of bell] (gen_do_2) {Generalized do-nothing\\M=2};

        \node[box, below=of bell] (cat) {Cat state\\M=3,4, J=2};
        \node[box, left=0.2cm of cat] (gen_trans_3) {Generalized transmit\\M=3};
        \node[box, right=0.2cm of cat] (gen_do_3) {Generalized do-nothing\\M=3};
        

        \draw[line] (transmit.south) -- ++(0,-0.4) -| (filter.north);
        \draw[line] (donothing.south) -- ++(0,-0.4) -| (filter.north);

        \draw[line] (filter.south) -- (bell.north);
        \draw[line] (filter.south) -- ++(0,-0.3) -| (gen_trans_2.north);
        \draw[line] (filter.south) -- ++(0,-0.3) -| (gen_do_2.north);

        \draw[line] (gen_trans_2.south) -- ++(0,-0.3) -| (gen_trans_3.north);
        \draw[line] (gen_do_2.south) -- ++(0,-0.3) -| (gen_do_3.north);
        \draw[line] (bell.south) -- (cat.north);

    \end{tikzpicture}%
    }
    \caption{AQT's chip optimal lookup workflow}
    \label{fig:aqt_workflow}
    \end{subfigure}
    \hfill
    \begin{subfigure}[t]{0.48\linewidth}
        \centering
    \resizebox{\textwidth}{!}{%
    \begin{tikzpicture}[
        node distance=0.8cm and 0.4cm,
        font=\footnotesize\sffamily,
        box/.style={
            draw, thick, rectangle, align=center,
            minimum height=1.1cm,
            text width=3.8cm, 
            inner sep=3pt
        },
        arrowline/.style={-Latex, thick},
        plainline/.style={thick}
    ]

        \node[box] (do_rest) {Do-nothing\\rest of assessment};
        \node[box, below=of do_rest] (gen_do_2) {Generalized \textit{do-nothing} M=2\\full assessment};
        \node[box, below=of gen_do_2] (gen_do_3) {Generalized \textit{do-nothing} M=3\\full assessment};

        \node[box, left=of do_rest] (trans_rest) {Transmit\\rest of assessment};
        \node[box, below=of trans_rest] (gen_trans_2) {Generalized \textit{transmit} M=2\\full assessment};
        \node[box, below=of gen_trans_2] (gen_trans_3) {Generalized \textit{transmit} M=3\\full assessment};

        \node[box, right=of do_rest] (bell_state) {Bell-state transfer\\full assessment};
        \node[box, below=of bell_state] (cat_3) {Cat state M=3 J=2\\full assessment};
        \node[box, below=of cat_3] (cat_4) {Cat state M=4 J=2\\full assessment};

        \coordinate[above=1.8cm of do_rest] (top_center);
        
        \node[box, left=0.2cm of top_center] (trans_c2c) {Transmit\\c2c};
        \node[box, right=0.2cm of top_center] (do_c2c) {Do-nothing\\c2c};


        \draw[arrowline] (trans_c2c) -- (do_c2c);

        \coordinate (bus_height) at ($(trans_c2c.south)!0.5!(do_rest.north)$);

        \draw[plainline] (do_c2c.south) -- (do_c2c.south |- bus_height);

        \draw[plainline] (trans_rest.north |- bus_height) -- (bell_state.north |- bus_height);

        \draw[arrowline] (trans_rest.north |- bus_height) -- (trans_rest.north);
        \draw[arrowline] (do_rest.north |- bus_height) -- (do_rest.north);
        \draw[arrowline] (bell_state.north |- bus_height) -- (bell_state.north);

        \draw[arrowline] (trans_rest) -- (gen_trans_2);
        \draw[arrowline] (gen_trans_2) -- (gen_trans_3);
        \draw[arrowline] (do_rest) -- (gen_do_2);
        \draw[arrowline] (gen_do_2) -- (gen_do_3);
        \draw[arrowline] (bell_state) -- (cat_3);
        \draw[arrowline] (cat_3) -- (cat_4);

    \end{tikzpicture}%
    }
    \caption{The optimal lookup workflow for IBM processors. An arrow $A \rightarrow B$ indicates that procedure B is executed only on sub-chips that successfully passed procedure A.}
    \label{fig:ibm_workflow}
    \end{subfigure}
    
    \caption{The optimal lookup workflows for AQT (left) and IBM (right)}
    \label{fig:workflows_figure}
\end{figure}

The initial strategy for assessing the AQT device involved utilizing all 12 qubits to explore their performance. However, preliminary data indicated significantly sub-optimal performance, with certain layouts exhibiting fidelities approaching 50\%. Consequently, we revised the workflow to first identify and exclude low-performing qubits, subsequently proceeding with the validation on the remaining subset, specifically the best eight out of twelve available qubits. The process of optimal lookup workflow was as follows:
\begin{enumerate}
    \item We execute \textit{transmit} and \textit{do-nothing} on all twelve qubits, based on the fidelities outcomes we identify the four lowest-performing qubits. For consistency between different days we always removed exactly four qubits
    \item We perform the remaining protocols - \textit{bell-state transfer}, \textit{generalized transmit} and \textit{generalized do-nothing} - exclusively on the reduced chip layout, altering the general protocols parameters as shown in Figure~\ref{fig:aqt_workflow} 
\end{enumerate}
While the \textit{cat state} protocol was initially planned to take part in this research, budget constraints forced us to remove it from the execution plans on AQT's computer. Consequently, \textit{cat state} protocol results are presented only for Fez quantum computer and do not take part in the comparison chapters.
The specific selection of excluded qubits on step 1 is determined by isolating the four lowest-performing qubits and evaluating the \textit{transmit} and \textit{do-nothing} fidelities with the remaining sub-set. If all the qubits in this configuration meet the required fidelity threshold, the layout is designated as the optimal sub-chip and the workflow proceeds to step 2. This adopted workflow is illustrated in Figure~\ref{fig:aqt_workflow}.

The adopted workflow for AQT's quantum computer cleaned the results and allowed a better assessment of the similarities and differences between the AQT's hardware and IBM's. The optimal lookup workflow of the IBM's computer is different because of a significant difference in qubit count. Instead of single-qubit granularity, IBM's assessment is done on rectangular sub-chips of 12 qubits each, a rectangle is always tested as one full unit. The full assessment stages were defined in the earlier work (\cite{Bench2_arxiv}), each stage excludes rectangles according to a defined fidelity threshold. There are three stages that compose the full assessment, each stage defines the tested inner paths in the sub-chip. The full assessment stages are as follows:

\begin{figure}[h]
    \centering
    \begin{tikzpicture}[
    x=0.95cm, y=0.8cm, 
    nnode/.style={circle, fill=myblue, text=white, inner sep=0pt, minimum size=22pt, font=\sffamily\small\bfseries},
    edge/.style={line width=4.5pt, edgegrey},
    pedge/.style={line width=4.5pt, mypurple}
]

\draw[edge] (3, 0) -- (10, 0);       
\draw[edge] (10, 0) -- (11, 0);     
\draw[edge] (3, -2) -- (11, -2);     
\draw[edge] (5, -4) -- (9, -4);      

\draw[edge] (3, 0) -- (3, -2);       
\draw[edge] (7, 0) -- (7, -2);       
\draw[edge] (11, 0) -- (11, -2);     
\draw[edge] (5, -2) -- (5, -4);      
\draw[edge] (9, -2) -- (9, -4);      

\foreach \x in {3,...,11} { 
    \node[nnode] at (\x, 0) {\x}; 
}

\node[nnode] at (3, -1) {16};
\node[nnode] at (7, -1) {17};
\node[nnode] at (11, -1) {18};

\foreach \x in {3,...,11} { 
    \pgfmathsetmacro{\val}{int(\x+20)} 
    \node[nnode] at (\x, -2) {\val}; 
}

\node[nnode] at (5, -3) {37};
\node[nnode] at (9, -3) {38};

\foreach \x in {5,...,9} { 
    \pgfmathsetmacro{\val}{int(\x+40)} 
    \node[nnode] at (\x, -4) {\val}; 
}

\node[font=\sffamily\Huge] at (5, -1) {1};
\node[font=\sffamily\Huge] at (9, -1) {2};
\node[font=\sffamily\Huge] at (7, -3) {5};

\end{tikzpicture}
    \caption{A close-up view of the qubit connectivity map for the IBM Heron-r2 architecture, illustrating the physical layout and grouping of the numbered rectangular sub-chips (e.g., 1, 2, and 5).}
    \label{fig:fez_qubit_map_sample}
\end{figure}
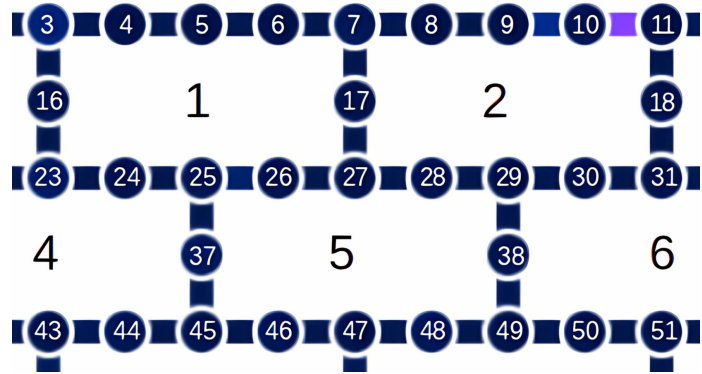
\begin{enumerate}
    \item \textbf{c2c (Corner-to-Corner)} - in this stage, we evaluate the internal paths of the sub-chip that connect qubits located at opposite corners of the rectangle. As an example figure~\ref{fig:fez_qubit_map_sample} shows two such paths: $3 \rightarrow 23 \rightarrow 27$ and $3 \rightarrow 7\rightarrow 27$. Each rectangle has eight paths that go from corner to corner
    \item \textbf{M-L (Maximal Lengths):} This stage evaluates all internal paths of the maximal length, which is seven qubits in this architecture. For example, in Figure~\ref{fig:fez_qubit_map_sample}, this includes the two paths starting at qubit 4 and ending at qubit 26 in rectangle 1. Each rectangle features 24 such layouts
    \item \textbf{A-L (All Lengths):} This final stage encompasses all minimal paths between every possible pair of qubits within the rectangle. Each rectangle contains a total of 144 such paths
\end{enumerate}

The fidelity threshold for each rectangle is defined per protocol, a rectangle is considered to pass the threshold if the minimum of the fidelities across all the inner paths pass the defined quantumness threshold. The optimal lookup workflow for assessing IBM's computer is shown in Figure~\ref{fig:ibm_workflow} and is composed of the following steps:
\begin{enumerate}
    \item First, we execute two filtration steps - the first step is \textit{transmit} c2c, removing the sub-chips that failed to pass the threshold. On the remaining rectangles sub-set we execute \textit{do-nothing} c2c stage, and again exclude the sub-chips that failed
    \item On the sub-chips that passed the previous stage we run the rest of \textit{transmit} and \textit{do-nothing} full assessment as well as \textit{bell-state transfer} full assessment
    \item With all rectangles that passed \textit{transmit} full assessment we proceed to \textit{generalized transmit} protocol full assessment with $M=2$. Then with successful rectangles proceed to full assessment of \textit{generalized transmit} with $M=3$. In the same way after the \textit{do-nothing} full assessment comes \textit{generalized do-nothing} with $M=2$ and then $M=3$. Originally, after the full assessment of \textit{bell-state transfer}, the experimental design included the execution of the full assessment of the \textit{cat state} protocol with $J=2$ and $M=3$ and then $M=4$. While this subsequent phase was generally omitted due to the budgetary constraints detailed in Section~\ref{sec:limitations_and_malfunctions}, we successfully executed the \textit{cat state} protocol on the IBM Fez system and present those results to demonstrate the complete, intended optimal lookup workflow.
\end{enumerate}
For more information and a more detailed explanation of IBM's chips assessment process, see \cite{Bench2_arxiv}.

\section{The Tested Hardware}\label{sec:hardware_intro}
\subsection{AQT's IBEX Q1}\label{sec:ibex_results_section}
We briefly present the quantum computer IBEX Q1, of Alpine Quantum Technologies (AQT), located in Austria. This quantum computer has 12 qubits with all-to-all connectivity, illustrated in Figure~\ref{fig:ibex_qubit_map}. Each qubit is assigned a number from 0 to 11. 

At the time of the experiments, AQT had a public execution window for the IBEX Q1 computer once a week. To prevent temporal inconsistency while still allowing us to correctly apply our proposed workflow, we performed the whole assessment over a single day, executing the first two selection stages (\textit{transmit} and \textit{do-nothing}) in the morning and after analysis of their outputs we execute the rest of the workflow. We performed this optimal lookup workflow on two separate days, 13th and 25th of August 2025. Supplemental Section~\ref{S-sec:ibex_results_section} presents the full results from the latter date, 25th of August 2025, this set of results is later used in the comparison section~\ref{sec:comparison_section}. Notably, on this day the performance of the IBEX Q1 computer was better than on the first day, 13th of Aug, the results of which are presented in the Supplemental Appendix~\ref{S-sec:ibex_13Aug_workflow}. Although the overall QPU performance and our qubit selection process were better on August 25th, the August 13th execution notably succeeded in identifying an optimal sub-chip for the \textit{generalized transmit} (M=2) protocol. It is worth noting that both of the tested IBM QPUs failed to produce an optimal rectangular sub-chip for this specific protocol.


\begin{figure}[H]
    \centering

    \begin{subfigure}[t]{0.48\linewidth}
        \centering
        \includegraphics[width=1\linewidth]{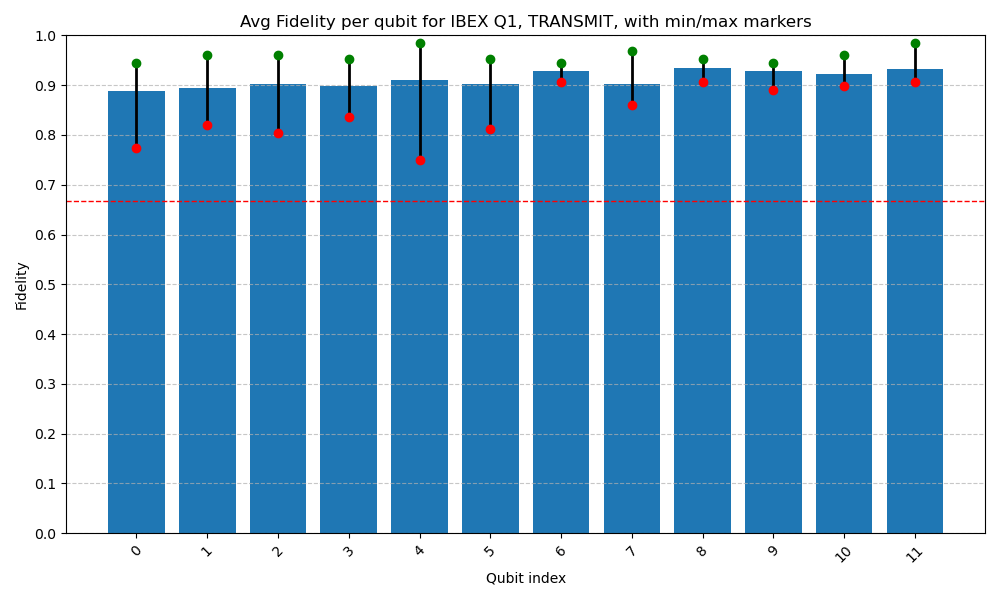}
    \caption{Transmit protocol on all 12 qubits, all qubits passed this stage}
    \label{fig:ibex_transmit_sixth_all_qubits}
    \end{subfigure}
    \hfill
    \begin{subfigure}[t]{0.48\linewidth}
        \centering
        \includegraphics[width=1\linewidth]{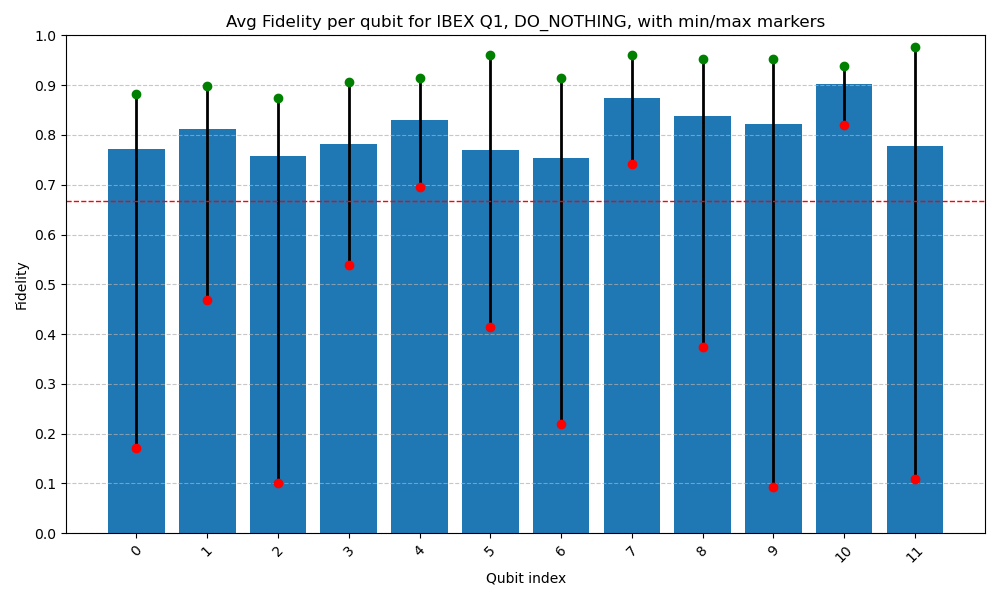}
        \caption{\textit{Do-nothing} protocol on all 12 qubits}
        \label{fig:ibex_do_nothing_sixth_all_qubits}
    \end{subfigure}

    \caption{The first two selection stages. This chart presents the mean, min and max fidelity as function of the measured qubit}
    \label{fig:ibex_sixth_first_selection_results}    
\end{figure}

To demonstrate the experimental methodology we present in Figures~\ref{fig:ibex_sixth_first_selection_results} the first stages of the data collection process. These figures show the first two experiments executed on the IBEX Q1 computer in this assessment. It is apparent that the chip showed optimal results on the \textit{transmit} protocol in Figure~\ref{fig:ibex_transmit_sixth_all_qubits}. In this experiment, we applied the \textit{transmit} protocol from every two qubits on the circuit bidirectionally, allowing us to assess the ability of each qubit to receive a state via swap gates from every other qubit in the chip.
On the \textit{do-nothing} protocol experiment, Figure~\ref{fig:ibex_do_nothing_sixth_all_qubits}, a sub-optimal performance is present in some of the qubits. Similar to the \textit{transmit} protocol, this experiment checks the ability of each qubit to perform a state swap with every other qubit and then receive that state back via another swap gate. 

\begin{figure}[H]
\centering
    \includegraphics[width=1\linewidth]{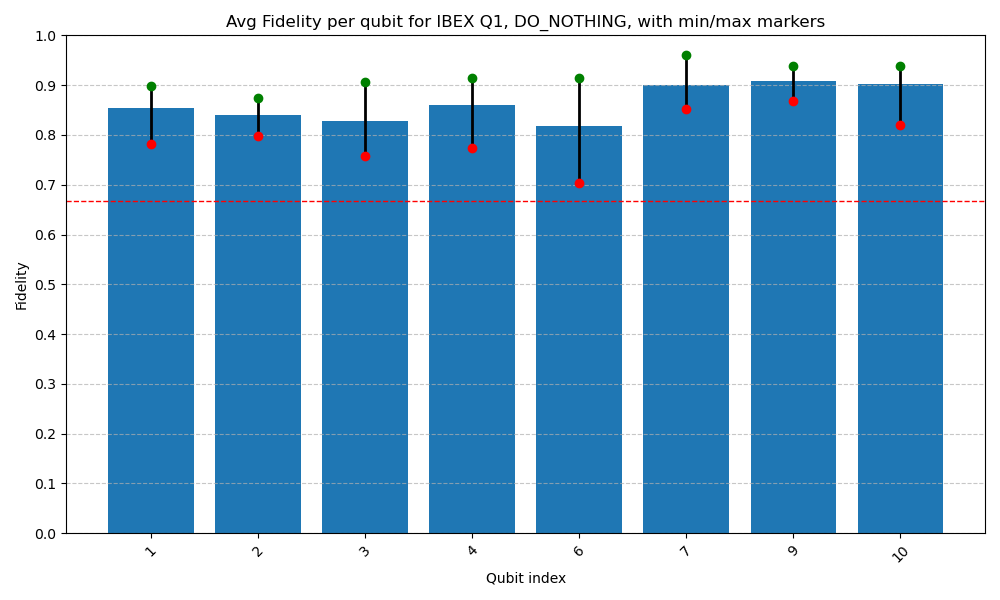}
    \caption{\textit{Do-nothing} protocol on all qubits except qubits 0, 5, 8 and 11 who were excluded due to poor performance}
    \label{fig:ibex_do_nothing_sixth_wo_0_5_8_11}  
\end{figure}

The selection process for identifying under-performing qubits was conducted through a pair-wise analysis rather than assessing each qubit in isolation. By evaluating the performance of the pairs, we were able to distinguish between qubits that were inherently faulty and those who only appeared sub-optimal due to interaction with a failing neighbor. This approach revealed that qubits 2, 6 and 9, for instance, remained functional despite initial aggregated data suggesting otherwise. Conversely, the analysis identified that qubits 0, 5, 8 and 11 were consistently involved in the lowest-performing pairs regardless of their partner. These four qubits were excluded and the results presented in Figure~\ref{fig:ibex_do_nothing_sixth_wo_0_5_8_11} were derived by re-aggregating the original experimental data after digitally filtering out this lowest-performing qubits set.



We see that removing these four qubits produces a truly quantum reduced chip, thereby we continue the rest of the optimal lookup workflow with the current optimal sub-set of qubits = $\{1,2,3,4,6,7,9,10\}$. As said, the full results for this assessment of IBEX Q1 are presented in the Supplemental Section~\ref{S-sec:ibex_results_section}.

\subsection{IBM's Hardware}\label{sec:Brisbane_results_section}
IBM's quantum computer, Brisbane, which belongs to an older series, Eagle-r3, serves in this work as the baseline reference for the two other tested QCs. The Brisbane chip is composed of 127 qubits, arranged in rectangles of 12 qubits each, as shown in Figure~\ref{fig:eagle_map}.  We assigned a number to each rectangle, treating it as a single independent unit. The granularity of our assessment of IBM's chips is solely on those rectangular units; thus, the results in Supplemental Section~\ref{S-sec:Brisbane_results} present fidelity as a function of the rectangle index in each protocol.

The last evaluated QC is IBM's Heron-r2 QC named Fez. This chip is composed of 156 qubits, arranged in a rectangular lattice of twelve-qubit sub-chips (similar to Brisbane), as seen in Figure~\ref{fig:fez_qubit_map}. The full results from the Fez QPU evaluation are presented in Supplemental Section~\ref{S-sec:fez_results_section} and Supplemental Appendix~\ref{S-sec:first_assessment_stages}.

To illustrate the experimental methodology on IBM's hardware, we present in this section results of the first step of the optimal lookup workflow on Brisbane. The rectangle selection process follows the workflow presented in Section~\ref{sec:optimal_lookup_workflow}, i.e., we run \textit{transmit} and do nothing c2c stage first, and only the rectangles that pass those two stages will proceed with the optimal lookup workflow as depicted in Figure~\ref {fig:ibm_workflow}.

As said, the first two stages of our optimal lookup workflow are the c2c stage of \textit{transmit} and then of \textit{do-nothing}. Such a selection process allows a fair comparison to AQT's chip, as its selection process is similar.

\begin{figure}[H]
    \centering

    \begin{subfigure}[t]{0.48\linewidth}
        \centering
        \includegraphics[width=\linewidth]{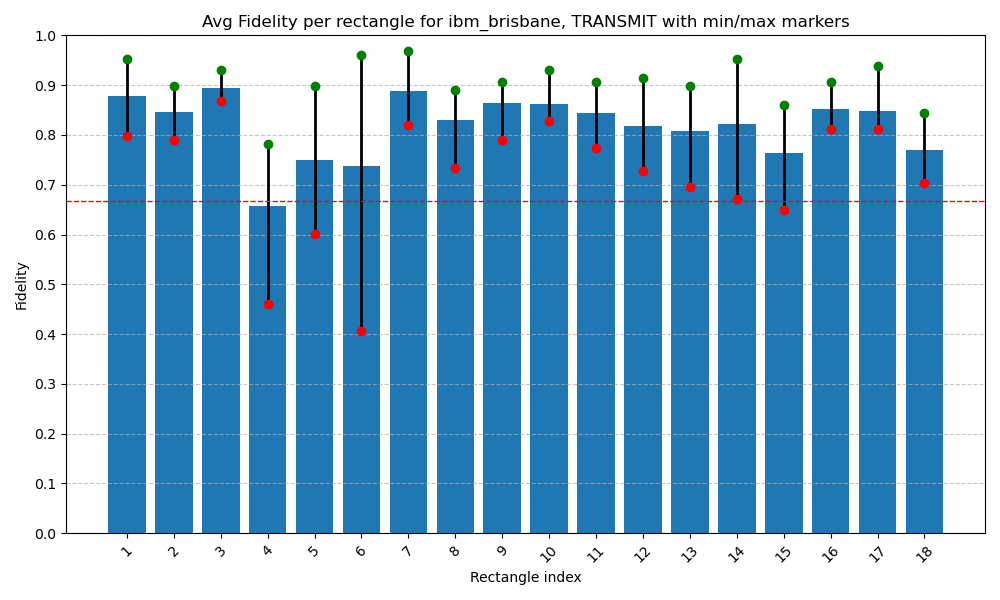}
    \caption{22 Aug 2025: \textit{Transmit}, c2c on all 18 rectangles of Brisbane}
    \label{fig:brisbane_transmit_c2c}    
    \end{subfigure}
    \hfill
    \begin{subfigure}[t]{0.48\linewidth}
        \centering
        \includegraphics[width=\linewidth]{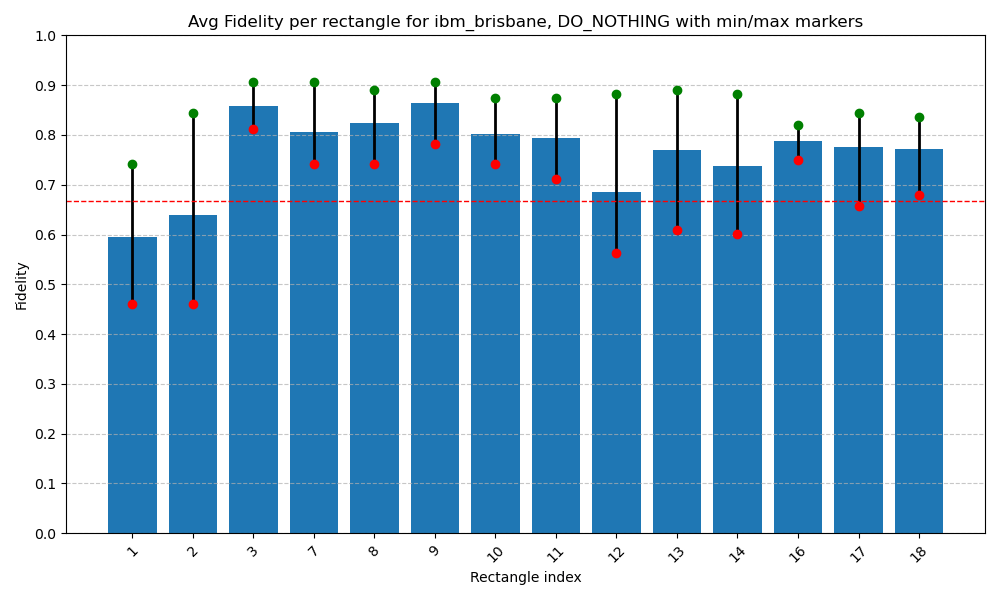}
        \caption{22 Aug 2025: \textit{Do-nothing} c2c on all rectangles except 4, 5, 6, 15, which failed \textit{transmit} c2c on 22 Aug 2025 (Figure~\ref{fig:brisbane_transmit_c2c})}
        \label{fig:Brisbane_do_nothing_c2c}  
    \end{subfigure}
    \caption{The results of the first selection stages, c2c of \textit{transmit} and \textit{do-nothing}, on Brisbane}
    \label{fig:brisbane_first_selection_stages}
\end{figure}
Figure~\ref{fig:brisbane_first_selection_stages} shows the results of the first selection stages on Brisbane. On the left, the \textit{transmit} protocol c2c stage is performed over all 18 rectangles of Brisbane. Rectangles 4, 5, 6 and 15 failed to provide above-threshold performance, as their minimum achieved fidelity (marked by the red dot on their bar) is below $\frac{2}{3}$, which is the quantumness fidelity threshold for this protocol. The right panel of Figure~\ref{fig:brisbane_first_selection_stages} presents the results of the \textit{do-nothing} protocol c2c stage over all the rectangles except 4, 5, 6 and 15. In the \textit{do-nothing} selection stage rectangles 1, 2, 12, 13, 14 and 17 have failed as well, providing the final sub-set of rectangles that will be tested in the next stages of the optimal lookup workflow. The sub-set of optimal rectangles is - \{3, 7, 8, 9, 10, 11, 16, 18\}. The rest of the results for this assessment process of Brisbane can be found in the Supplemental Section~\ref{S-sec:Brisbane_results}.

\begin{figure}[t]
    \centering
    \begin{tikzpicture}[
    x=0.9cm, y=0.8cm, 
    nnode/.style={circle, fill=myblue, text=white, inner sep=0pt, minimum size=20pt, font=\sffamily\small\bfseries},
    edge/.style={line width=4.5pt, edgegrey}
]

    \draw[edge] (0, 0) -- (13, 0);
    \draw[edge] (0, -2) -- (14, -2);
    \draw[edge] (0, -4) -- (14, -4);
    \draw[edge] (0, -6) -- (14, -6);
    \draw[edge] (0, -8) -- (14, -8);
    \draw[edge] (0, -10) -- (14, -10);
    \draw[edge] (1, -12) -- (14, -12);
    
    \foreach \x in {0, 4, 8, 12} {
        \draw[edge] (\x, 0) -- (\x, -2);
        \draw[edge] (\x, -4) -- (\x, -6);
        \draw[edge] (\x, -8) -- (\x, -10);
    }
    \foreach \x in {2, 6, 10, 14} {
        \draw[edge] (\x, -2) -- (\x, -4);
        \draw[edge] (\x, -6) -- (\x, -8);
        \draw[edge] (\x, -10) -- (\x, -12);
    }
    
    \foreach \x in {0,...,13} { \node[nnode] at (\x, 0) {\x}; }
    \foreach \x in {0,...,14} { \pgfmathsetmacro{\val}{int(\x+18)} \node[nnode] at (\x, -2) {\val}; }
    \foreach \x in {0,...,14} { \pgfmathsetmacro{\val}{int(\x+37)} \node[nnode] at (\x, -4) {\val}; }
    \foreach \x in {0,...,14} { \pgfmathsetmacro{\val}{int(\x+56)} \node[nnode] at (\x, -6) {\val}; }
    \foreach \x in {0,...,14} { \pgfmathsetmacro{\val}{int(\x+75)} \node[nnode] at (\x, -8) {\val}; }
    \foreach \x in {0,...,14} { \pgfmathsetmacro{\val}{int(\x+94)} \node[nnode] at (\x, -10) {\val}; }
    \foreach \x in {1,...,14} { \pgfmathsetmacro{\val}{int(\x+112)} \node[nnode] at (\x, -12) {\val}; }
    
    \foreach \x/\val in {0/14, 4/15, 8/16, 12/17} { \node[nnode] at (\x, -1) {\val}; }
    \foreach \x/\val in {2/33, 6/34, 10/35, 14/36} { \node[nnode] at (\x, -3) {\val}; }
    \foreach \x/\val in {0/52, 4/53, 8/54, 12/55} { \node[nnode] at (\x, -5) {\val}; }
    \foreach \x/\val in {2/71, 6/72, 10/73, 14/74} { \node[nnode] at (\x, -7) {\val}; }
    \foreach \x/\val in {0/90, 4/91, 8/92, 12/93} { \node[nnode] at (\x, -9) {\val}; }
    \foreach \x/\val in {2/109, 6/110, 10/111, 14/112} { \node[nnode] at (\x, -11) {\val}; }
    
    \node[nnode] at (11, 0) {11};
    \node[nnode] at (6, -2) {24};
    \node[nnode] at (12, -6) {68};
    \node[nnode] at (2, -12) {114};
    
    \foreach \y/\startval in {-1/1, -5/7, -9/13} {
        \foreach \x [count=\xi from 0] in {2, 6, 10} {
            \pgfmathsetmacro{\val}{int(\startval+\xi)}
            \node[font=\sffamily\Huge] at (\x, \y) {\val};
        }
    }
    \foreach \y/\startval in {-3/4, -7/10, -11/16} {
        \foreach \x [count=\xi from 0] in {4, 8, 12} {
            \pgfmathsetmacro{\val}{int(\startval+\xi)}
            \node[font=\sffamily\Huge] at (\x, \y) {\val};
        }
    }
    
    \end{tikzpicture}
    \caption{Eagle-r3 series qubits map and rectangle indexes}
    \label{fig:eagle_map}
\end{figure}


\section{Results and Comparison}\label{sec:comparison_section}

To compare the AQT and IBM architectures, we place a primary emphasis on the minimal fidelity metric obtained for each Quantum Processing Unit (QPU) using our proposed optimal lookup workflow. Table~\ref{tab:comparing_min_fidelity} details these fidelity metrics alongside the specific parameters yielded by our experimental procedure for each protocol:
\begin{enumerate}
\item \textbf{For AQT} - The number of qubits in the optimal reduced chip and its minimal achieved fidelity.
\item \textbf{For IBM} - The number of optimal sub-chips, alongside the average minimal fidelity over those sub-chips.
\end{enumerate}
The data is presented in Table~\ref{tab:comparing_min_fidelity}. In this table, a hyphen (-) indicates that the chip was tested but failed to produce an optimal reduced chip (for AQT) or failed to produce optimal sub-chips (for IBM). Empty cells indicate that the experiment was not conducted. For AQT and Brisbane, these omissions were due to budgetary and availability constraints as detailed in Section~\ref{sec:limitations_and_malfunctions}, whereas for Fez, they resulted from the exclusion of all sub-chips in earlier selection stages.
\begin{table}[h]
    \centering
    \resizebox{\linewidth}{!}{
    \begin{tabular}{c|cc|cc|cc}
        \toprule
         \multirow{2}{*}{\textbf{Protocol}} & \multicolumn{2}{c}{\textbf{IBEX Q1}} & \multicolumn{2}{c}{\textbf{Eagle-r3 Brisbane}} & \multicolumn{2}{c}{\textbf{Heron-r2 Fez}}  \\


         & \textbf{Sub-Chip Size} & \textbf{Min F.} & \textbf{Sub-Chips} & \textbf{Avg Min F.} & \textbf{Sub-Chips} & \textbf{Avg Min F.} \\
         \midrule
         \textit{Transmit}            & 8 & 0.75     & 7 & 0.806  & 10 & 0.82 \\
        \textit{Generalized transmit} M=2    & - & -        & - & -     & - & - \\
        \textit{Generalized transmit} M=3    & - & -        &   &       &   &   \\
         \textit{Do-nothing}         & 8 & 0.703    & 5 & 0.745 & 10 & 0.785 \\
        \textit{generalized do-nothing} M=2  & - & -        &   &       & 10 & 0.654 \\
         \textit{generalized do-nothing} M=3  & - & -        &   &       & 10 & 0.533 \\
         \textit{Bell-state transfer} & 5 & 0.5      &   &       & 9 & 0.696 \\
         \textit{Cat state} J=2,M=3   &   &          &   &       & 7 & 0.681 \\
         \textit{Cat state} J=2,M=4   &   &          &   &       & - & - \\
         \bottomrule
    \end{tabular}
    }
    \caption{Comparative performance of AQT's optimal reduced chip versus IBM's optimal sub-chips. A hyphen (`-') indicates that the experiment was conducted but yielded no successful reduced chip (for AQT) or no sub-chips passed the threshold (for IBM). Empty cells indicate the experiment was not conducted due to budget constraints (AQT) or failure in prior stages (IBM)}
    \label{tab:comparing_min_fidelity}
\end{table}

A comparative analysis of the baseline protocols — \textit{transmit} and \textit{do-nothing} — reveals comparable performance capabilities between the AQT IBEX Q1 and IBM Brisbane systems. It is important to note that the results presented here belong to a new version of the Brisbane quantum computer, as depicted in~\cite{Bench2_arxiv} under the name Modified Brisbane, this version of Brisbane had significant improvements over the older version. Specifically, in this study, the minimal fidelity observed on IBEX Q1 was commensurate with the minimal fidelity of the optimal sub-chips on modified Brisbane. It is crucial to note that IBEX achieved this parity while offering superior connectivity (all-to-all), in stark contrast to the nearest-neighbor constraints in the modified Brisbane quantum computer. However, this architectural flexibility comes with an inherent trade-off: the superconducting system allows for the parallel execution of two-qubit gates, a capability precluded in the IBEX system due to its reliance on a centralized phonon mode. Given that only a limited number of sub-chips out of the 18 tested sub-chips on modified Brisbane met the fidelity threshold, this comparison indicates that IBEX Q1 delivers better performance than the Brisbane system when architectural flexibility is considered. It is apparent that IBM's Fez chip shows better results than the AQT chip, both in fidelity performance and in number of usable optimal regions within the chip. However, this must be contextualized: despite the Heron processor's 21 rectangular sub-chips in total, fewer than half of them successfully passed our protocols. In the generalized protocols, both chips had a steep decrease in performance, with AQT's chip failing to produce optimal results with every generalized protocol, and IBM's chip producing optimal sub-chips only in \textit{generalized do-nothing} M=2,3 and \textit{cat state} J=2, M=3.
Notably, while the Fez system failed to yield any optimal sub-chips for the \textit{generalized transmit} protocol, it consistently demonstrated optimal performance in the \textit{generalized do-nothing} protocol for both values of M. This divergence in results was unexpected, and the underlying cause for this performance discrepancy remains to be determined.
Although IBM's chip appears superior to IBEX, the qubit connectivity constraints of AQT's chip are a strong advantage. Our benchmarking method expresses this advantage over the reduced swap distance, as explained in the protocols Section~\ref{sec:protocols_explanations}, the major difference between the two versions of each protocol for IBM and AQT is the swap distance. For IBM, it can vary from one to six, depending on the protocol, and for AQT it is fixed to one (to the extent of a centralized phonon). This fact shows a great advantage of the benchmarking via protocols method, it allows an abstraction of such differences, focusing solely on quantumness advantage qualities.

\section{Discussion}\label{sec:discussion}
The comparative nature of this study, involving two distinct quantum architectures (Superconducting and Trapped-Ion), introduced specific logistical and economic challenges that shaped our experimental methodology. We also discuss in Section~\ref{sec:benchmarking_logical_qpu} a third quantum architecture, atom arrays, which may not be as good as the two above for benchmarking physical qubits but may currently be the optimal architecture for fault-tolerant quantum error correction codes (FT-QECC). To provide our intuition about the advantage of ``atom arrays technology and architecture'', we first clarify in Section~\ref{sec:benchmarking_modular_qpu} modular quantum computers based on various leading technologies.

\subsection{Limitations and Malfunctions}\label{sec:limitations_and_malfunctions}
Benchmarking across the AQT IBEX Q1 and IBM platforms (Fez and Brisbane) was shaped by significant differences in budget constraints, temporal stability, and hardware accessibility. Due to AQT's higher cost-per-shot compared to IBM's execution-time pricing model, the ion-trap experiments required a lower sampling resolution and a highly optimized workflow to maximize information gain. Furthermore, data acquisition between the platforms was separated by a three-month gap, during which the AQT system exhibited severe temporal drift. This instability, combined with AQT's restricted access window of just one day per week, forced all AQT experiments to be executed in a rapid, continuous "sprint" to prevent the invalidation of earlier qubit selection stages. In contrast, while earlier IBM models experienced availability challenges, the cloud-queued IBM Fez enabled flexible, comprehensive data collection, with both architectures ultimately benefiting from testing windows that were free of major maintenance downtimes.

\subsection{Benchmarking Modular Quantum Computers with Protocols}\label{sec:benchmarking_modular_qpu}

While this work focused on the performance of IBM's individual 12-qubit rectangular sub-chips, the natural evolution of our benchmarking-via-protocols framework is to evaluate larger, integrated regions. A logical step is to test successful pairs of neighboring optimal rectangles as unified sub-chips - an extension we have previously demonstrated on the IBM Brisbane (Eagle-r3 series) and Kingston (Heron-r2 series) QPU~\cite{Bench2_arxiv}. 
Furthermore, expanding this strategy to include triplets, quadruplets and even larger modular configurations offers a promising path for future research, providing a more granular understanding of how ``quantumness'' is preserved as systems scale. 

Recent modular frameworks, such as architectures utilizing bivariate bicycle codes, highlight that managing error correction across these expanding configurations is critical for viable modular computing~\cite{yoder2025}. Similarly, in trapped-ion architectures, scaling is being realized through advanced Quantum Charge-Coupled Device (QCCD) layouts, such as race-track processors that rely on extensive ion shuttling to maintain connectivity across modular zones~\cite{Moses_2023}.

This trajectory is especially vital as the industry moves towards large-scale modularity, where evaluating the performance of multi-chip architectures represents a critical frontier. Those systems often rely on interconnects, such as recent demonstrations of meter-scale physical links between large quantum processors~\cite{sundaresan2026}, to link separate quantum processors, providing a unique challenge for state transfer and entanglement protocols as those presented in this work. Applying our binary benchmarking methodology to these inter-chip channels would allow for a quantitative assessment of how modular scaling impacts the overall ``quantumness'' of these systems. The fidelity bottlenecks inherent to these interconnects and shuttling operations underscore the potential advantages of neutral-atom arrays, whose dynamic and reconfigurable connectivity offers a native solution for scaling without relying heavily on these said interconnects and shuttling operations~\cite{Bluvstein_2023, xu2024constant}.


\subsection{Benchmarking Logical QPUs via Protocols}\label{sec:benchmarking_logical_qpu}

We present a short discussion on the challenges of FT-QECC and suggest a concrete hardware protocol to rigorously test fault-tolerance claims. As computing shifts to logical qubits, physical gate overhead becomes substantial. Our low-gate-count benchmarking via protocols strategy is ideally suited to characterize these early-stage architectures while minimizing algorithmic overhead.

FT-QECCs are the framework required to protect quantum information from the noise exhibited during different calculations, a critical prerequisite for scalable quantum computation. This protection is achieved by encoding information into logical qubits, using, for example, stabilizer formalisms such as Calderbank-Shor-Steane (CSS) codes~\cite{Calderbank_1996, steane_1996} and in particular Steane's [[7,1,3]] color code ~\cite{steane_1996, reichardt2005quantum}, alongside other relevant directions like scalable surface code architectures~\cite{google2023suppressing, google2025quantum}, quantum error mitigation~\cite{cai_2023}, and magic-state distillation protocols~\cite{meier_2013}. The Eastin-Knill theorem~\cite{Eastin_2009} demonstrates that no quantum error-correcting code can transversely implement a universal gate set using a single D-dimensional code. To bypass this limitation and achieve universality, the process commonly relies on magic states (e.g. $\ket T$), which replace the relevant gate (e.g. a T gate replaced by the state $\ket{T}$) on logical qubits. Because the hardware is noisy, the magic states cannot be prepared fault-tolerantly directly within a code block. To overcome the noise, the hardware executes the magic state distillation process~\cite{Bravyi_2005}, a resource-intensive purification process that turns several low-fidelity magic states into a single, purified high-fidelity magic state. The distilled magic state is then injected~\cite{gottesman1999demonstrating, knill2005} into the logical circuit to execute non-Clifford operations such as the T-gate needed for universality~\cite{Boykin_1999, BOYKIN2000101}. The use of merging (2D and/or 3D) codes allows transversal implementation of a universal set, but adds a need for teleportation between the codes, offering an alternative to the overhead of standard magic state distillation~\cite{bluvstein2026fault, scruby2025fault}.


Recent implementations of neutral-atom QPUs have demonstrated the realization of essential components for fault-tolerant architectures~\cite{Bluvstein_2023}. These systems utilize reconfigurable atom arrays to achieve dynamic connectivity, parallel transversal gates, and atom-loss detection for error decoding~\cite{zhao2026ultrahighratequantumerrorcorrection}. Recent proofs-of-concept have utilized this approach to execute logical circuits with mid-circuit error removal~\cite{bluvstein2026fault} and demonstrate logical magic state distillation~\cite{rodriguez_2025}. Recent works in the field have also presented successful demonstrations of gate injection on IBM's superconducting quantum computers~\cite{Kim_2026}, as well as the fault-tolerant execution of error-corrected algorithms on trapped-ion systems~\cite{perlin2026faulttolerantexecutionerrorcorrectedquantum}.

To utilize our rigorous benchmarking approach to test these emerging fault-tolerant capabilities, we propose a \textit{transmit} or \textit{do-nothing} protocol executed between two logical qubits, one for Alice and one for Bob. Implementing these simplest benchmarking methods ``via protocols" already challenges the fundamental building blocks of a logical QPU across a few critical steps. First, the system must initialize a state and apply a specific single-qubit rotation; Using Solovay-Kitaev synthesis  (relevant to FT-QECC is Bravyi-Haah calculation~\cite{bravyi_2012}), this rotation can be realized as a minimal non-trivial approximation requiring just three $T$-gates interspersed with standard Clifford operations. Second, the logical state must be swapped to the second logical qubit, which can be accomplished using a standard SWAP gate requiring three transversal CNOTs, or a more efficient polarization SWAP requiring only two transversal CNOTs (assuming the second logical qubit is in the $\ket{0}$ state). Finally, the protocol concludes by ``un-rotating" the state back into the $\ket{0}$ state using a sequence of $T^\dagger$-gates. By evaluating the fidelity of the final measurement and comparing it against the classical limit, this straightforward challenge provides a direct metric for a QPU's ability to maintain logical coherence through non-Clifford injection, transversal entanglement, and multi-block routing. Even the basic \textit{transmit} and \textit{do-nothing} protocols are highly challenging, as applying just three T-gates for Alice and three inverse gates for Bob demands a massive physical qubit overhead to prepare, distill, and inject six logical T-state ancillas, or a highly complex merge of two codes. Consequently, more advanced rotations~\cite{bravyi_2012} and protocols (see Section~\ref{sec:protocols_explanations}) will require significantly larger numbers of both logical and physical qubits. Far from a universal rotation, a single-qubit transmit or do-nothing with just a single T-gate for Alice and a single T-gate for Bob seems to be achievable with several architectures in the very near future.

\section{Conclusions}\label{sec:conclusions}
In conclusion, we have demonstrated the utility of Benchmarking via Protocols as a robust metrics producer for evaluating quantum performance across fundamentally different physical implementations of a quantum computer. Our comparative analysis highlights distinct architectural advantages: the AQT's IBEX Q1 system excels in the regular protocols, as the performance is similar and the connectivity map is much more flexible, while the Heron-r2 superconducting system demonstrates superior performance on most of the generalized protocols. Notably, while the IBM architectures exhibited significant generation-over-generation improvement, the effective computational size of these processors remains substantially smaller than their total physical qubit count. 

This work also highlights the critical importance of operational transparency in quantum hardware to achieve reliable and rigorous benchmarking, an area where both AQT and IBM excel. By granting researchers the granular control necessary to isolate and evaluate specific sub-regions, these providers actively support independent, objective assessments of quantum advantage. In addition, we would like to express our appreciation to AQT and IBM for offering accessible pricing models that facilitate academic research, as this financial accessibility is essential for conducting the extensive, protocol-based evaluations presented in this study on chips and many sub-chips. 

We recommend that quantum firmware developers adopt the Benchmarking via Protocols methodology to rigorously and intuitively demonstrate the quantumness and practical usability of their chips and optimal sub-chips. 

The results of this research also suggest several directions for future research on collections of physical qubits to further refine the Protocol-based Benchmarking methodology. A primary objective is the expansion of this framework to a wider array of available platforms and physical architecture types to continue to build a robust, architecture-agnostic ``common language'' for QC benchmarking. While this work primarily focuses on circuit-based architectures, future efforts could explore the flexibility of protocol-level abstraction when applied to no-circuit paradigms, such as Measurement-Based QC (MBQC)~\cite{PhysRevLett.86.5188, Lanyon_2013, Wei_2021}, Adiabatic Quantum Computing (AQC)~\cite{Albash_2018} and ``boson sampling''-type non-universal linear optical quantum computing (non universal LOQC)~\cite{aaronson_2010, Gard_2015}, to test the universality of our binary criteria.

A note added after our results became public (ArXiv 2603.27397 v1): Following a request by AQT personnel who suggested that potentially there might have been some problem with IBEX-Q1 in August 2025, we repeated three of our benchmarking experiments in April 2026. The results are provided in the Supplemental Appendix~\ref{S-sec:additional_data_ibex} and are indeed better than the results from August 2025.

\section{Acknowledgment}
We thank the Quantum Computing Consortium of the Israel Innovation Authority and Classiq Technologies LTD for financial support. This research project was partially supported by the Helen Diller Quantum Center at the Technion. We thank Ilana Frank Mor for reading parts of this manuscript and providing many useful comments. TM and YW thank Chen Mechel and Rotem Liss for their contribution to the very early stage of experimenting with teleportation and entanglement swapping, two benchmarking protocols later used in~\cite{Bench1_arxiv, Bench2_arxiv}.
T.M. thanks Yehuda Naveh (from Classiq Technologies) for a recent discussion leading to viewing the benchmarking of logical qubits via protocol research direction as becoming feasible these days.

\printbibliography

\end{document}